# Numerical verification of the existence of localization of the elastic energy for closely spaced rigid disks

S. I. Rakin

Siberian State University of transport

Russia, 630049, Novosibirsk, Dusy Kovalchuk street, 191

e-mail: www. rakinsi@ngs.ru

The problem of determining the thermomechanical characteristics of the system of closely spaced bodies considered by many authors, see [1-9] and reviews on the problem in [10, 11]. For scalar problems, such as problems of heat, electrostatic, etc., the localization effect was found [12-16] (so called Tamm shielding effect [16]), consists in the fact that in a system of closely spaced highly conductive bodies, most energy is localized in the region between adjacent bodies. We can assume that the localization energy is enough for a general property of closely spaced bodies. The aim is to study the effect of localization of energy in the closely spaced rigid bodies in an elastic medium.

## 1. Numerical verification of localization of the elastic energy

Keeping in mind the sustained interest in the problem of the behavior of systems of closely spaced elastic bodies (see [1-9] and references in [10, 11]), could have been expected to have experimental data on the distribution of local fields in such systems. However, as in the case of the problem of thermal conductivity [12], the author of such data could not be found. The reason seems to be the same as in the heat conduction problem - the difficulty of measuring fields in the narrow spaces between adjacent bodies. Therefore, as in [12] as a research tool was chosen numerical method. That is, a numerical solution of the problem of elasticity theory using the program ANSYS. For the elasticity problem of numerical calculations on ANSYS are highly accurate and can be regarded as the equivalent of the experiment.

Considered next plane problem of elasticity theory. There are two fairly hard disk placed in the relatively soft matrix linearly elastic and close to each other, see Fig. 1. Disks are moved as solids. The mutual displacement of the two drives on the plane can be represented as the sum of the following basic movements of disks relative to each other, "up-down", "left-right", "rotated on one side", "rotate in opposite directions."

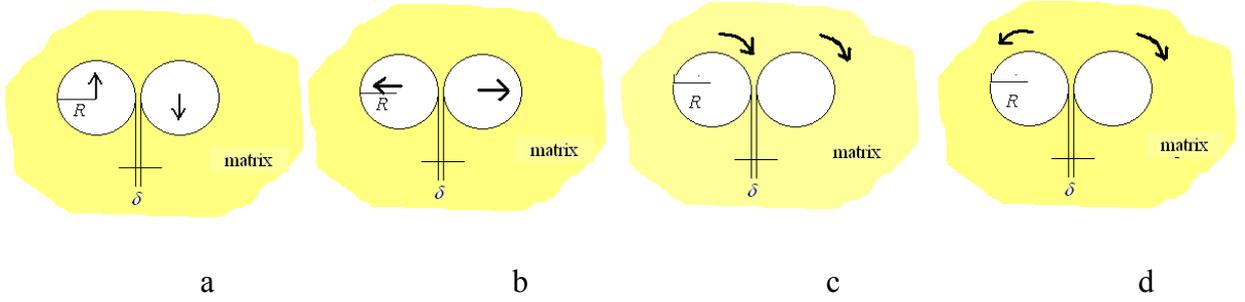

**Figure 1.** Location and movement of the disc-inclusion in the matrix

The calculations theoretically unlimited matrix replaced by a "sufficiently large" area. Namely, we consider disk of large radius $R_1 = 25$, at the center of the disk were placed symmetrically arranged hard disks of the radius $R_2 = 1$ at a distance $\delta$ from each other. On the border of a large disk displacement were set equal to zero. At the boundaries of disks - matrix assume that the conditions ideal contact. The characteristic size of the problem is the relative distance between the disks. For closely spaced disks $\frac{\delta}{R_1}$ is much less than 1. In the calculations, the range of variation is $\delta \in [10^{-4}; 10^{-2}]$. The elastic properties of the matrix materials and disks as follows: Young's modulus of the matrix $E_1 = 1$, the modulus of disks $E_2 = 1000$ (with such a difference of Young's modulus the disks behave as almost absolutely rigid body), Poisson's ratio $\nu = 0.3$.

For each of the basic movements of hard drives solved the problem of the plane theory of elasticity. Moving discs were taken of the same order with distance $\delta$. The displacement along the $Ox$ axis is denoted by $u$. The displacement along the $Oy$ axis is denoted by $\upsilon$.

1. Basic movement "up-down" (Fig. 1a): For one disk $u = 0$, $\upsilon = +\delta$. For another disc $u = 0$, $\upsilon = -\delta$.
2. Basic movement "left-right" (Fig. 1b): For one disk $u = -\delta$, $\upsilon = 0$. For another disc $u = +\delta$, $\upsilon = 0$.
3. Basic movement "rotation in the same direction" (Fig. 1c): On the border of the disks the tangential movement $u_\varphi = \delta$, the radial displacement $u_r = 0$.
4. Basic movement "rotation in different directions" (Fig.1d): On the border of one disk the tangential displacement constant $u_\varphi = \delta$, the radial displacement $u_r = 0$. For another disk the tangential displacement $u_\varphi = -\delta$, radial displacement $u_r = 0$.

In the heat conduction problem localization is most pronounced for the energy and less pronounced for the heat flow and gradients [12, 13]. In our calculations, we observed a similar situation: the localization is most pronounced for energy, and less pronounced for the strain and stress.

Localization of elastic energy illustrated with Figs 2,3,4,5. The calculation was performed for the distance between the disks $\delta = 0.01$. For the numerical characteristics of localization consider the elastic energy $E_{mat}$ of the whole matrix and elastic energy $E_{sh}$ in the neck between the discs (neck region was chosen in accordance with Figs 2,3,4,5). The ratio $\frac{E_{sh}}{E_{mat}}$ is the proportion of the total elastic energy of the matrix, concentrated in the neck between the disks (the size of the neck was taken $2/3R_2$).

Thus, the total energy of elastic deformation of the matrix and the elastic energy of the disks in the neck, for the 1st, 2nd and 3rd base movements closely spaced hard disks are practically the same. That is, for these basic movements takes place the asymptotic (as $\delta \to 0$) localization of the elastic deformation energy of the matrix.

**Table 1.** The share of total energy matrix entered between the discs in the neck

| $\dfrac{E_{sh}}{E_{mat}}$ | Rotation, different directions | "left-right" | Rotation, the same direction | "up-down" |
|---|---|---|---|---|
| $\delta = .18$ | 30% | 20% | 73% | 32% |
| $\delta = .08$ | 30% | 84% | 80% | 83% |
| $\delta = .04$ | 43% | 89% | 85% | 87% |
| $\delta = .02$ | 44% | 92% | 90% | 91% |
| $\delta = .01$ | 46% | 95% | 93% | 94% |
| $\delta = .001$ | 50% | 98% | 98% | 98% |
| $\delta = .0001$ | 50% | 99.7% | 99.6% | 99.7% |

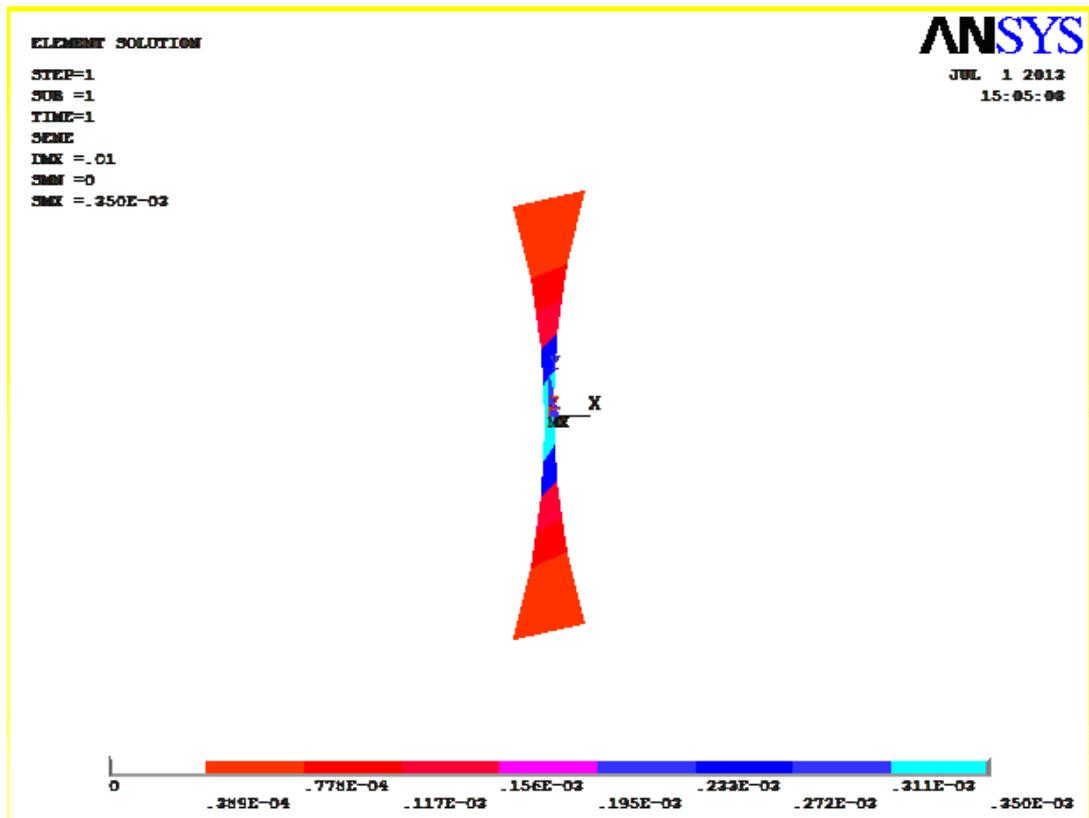

**Fig.2** The energy of the elastic deformation under disk moving "up-down"

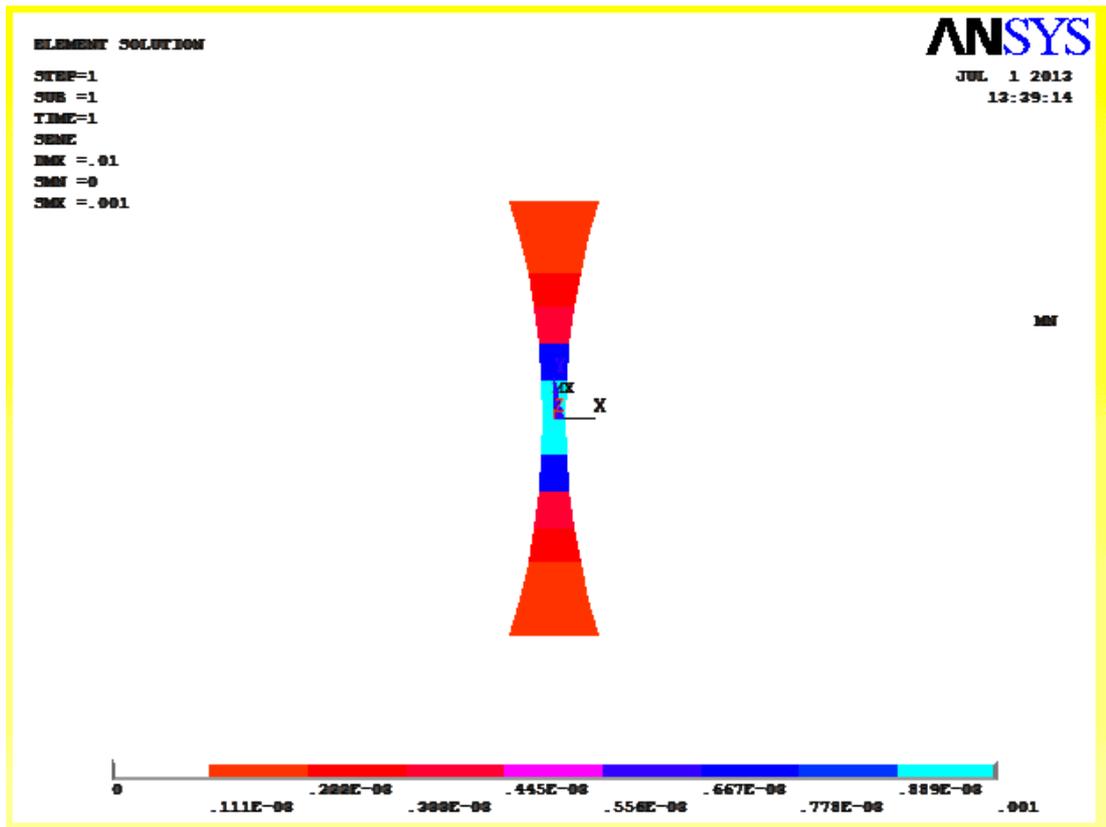

**Fig.3** The energy of the elastic deformation under the disk moving "left-right"

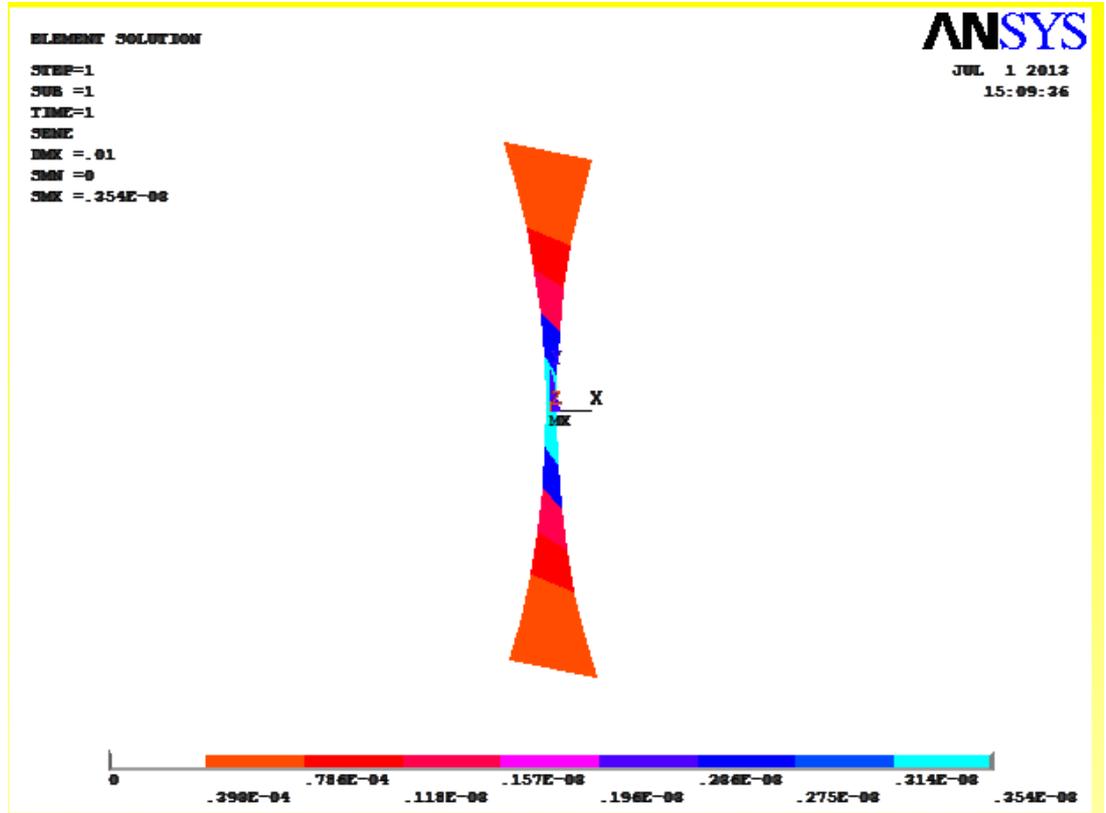

**Fig.4** The energy of the elastic deformation under the rotation disk in the same direction

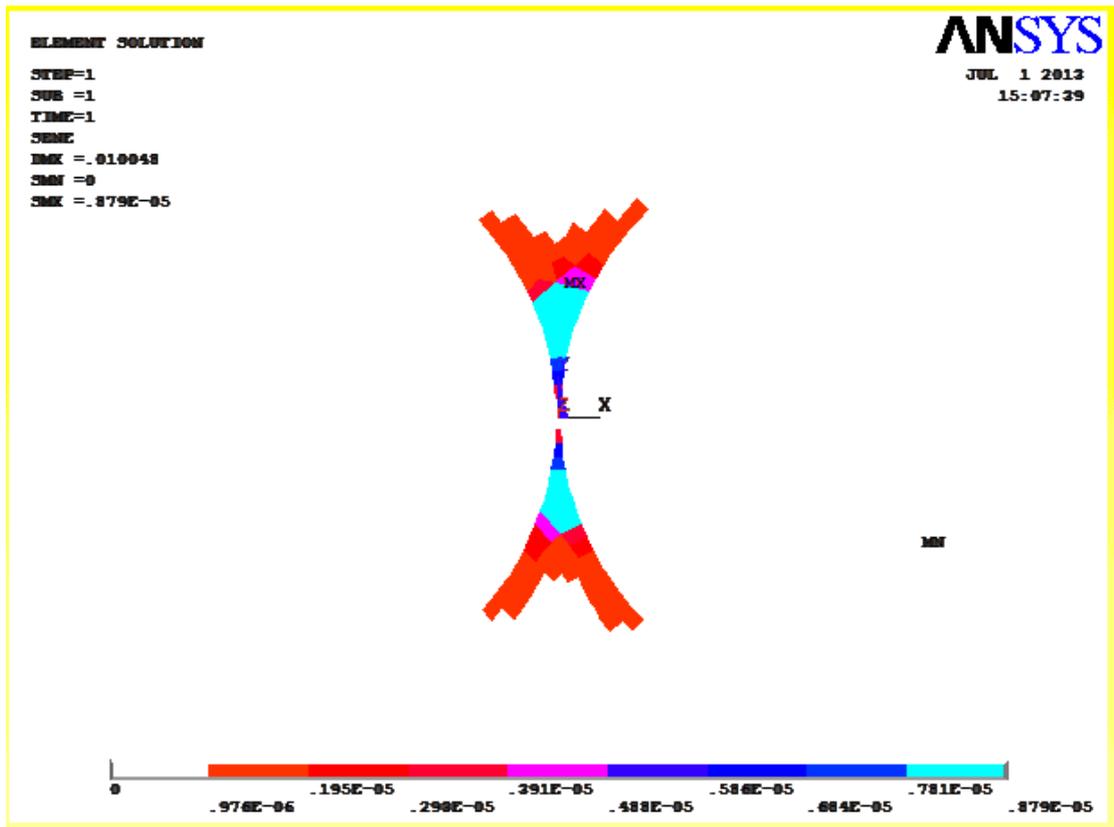

**Fig.5** The energy of the elastic deformation under the rotation disk in different directions

**Table2.** Calculated values for energy of basic movements

| Distance $\delta$ | "up-down" | "left-right" | One direction | Different directions | Different directions / "up-down" |
|---|---|---|---|---|---|
| $\delta = .01$ | 9.49e-3 | 13.21e-3 | 13,19e-3 | 1,77e-3 | 18.7% |
| $\delta = .001$ | 34,67e-5 | 49.08e-5 | 49.08e-5 | 1.85e-5 | 5.3% |
| $\delta = .0001$ | 214.93e-7 | 214.55e-7 | 215.11e-7 | 1.91e-7 | 0.9% |

For the 4th basic movements - "rotation drives in different directions," a significant portion of the energy is collected in the neck between the discs. However, as can be seen from Table 1, in the neck it is collected about half elastic energy, the second half of the energy is out of the neck. That is, in this case, there is concentration of energy but no localization (as defined in [13], the concentration of the local field in a region is to achieve a relatively high field in this region, and localization is collection of almost all the fields in this area).

Calculated values of the energy of deformation of the matrix (see Table 2) show that for the 4th base movement, the energy is small compared with the energies corresponding to the 1st, 2nd and 3rd base movements (see the last column of Table 2). In this connection it can be neglected.

In the calculations presented in Table 2, the basic movements of movements in the "left-right", "up-down" assumed to be equal $\delta$. When rotating "one-way", "in different directions" is taken to be moving $\delta$.

## 2. Formula for calculating the elastic energy of matrix by moving discs

The linearity of the energy problem is a quadratic function of the variables $u, \upsilon, \omega$. We need to calculate the coefficients of the quadratic form.

*Remark.* In the problem of electrostatics and thermal conductivity [19], the energy is a quadratic function of the difference of potentials or temperatures of disks $t$: $E = 1/2 ct^2$ where the coefficient $c$ is the harmonic (electrostatic) capacity. We need to calculate the capacity of analogues for the elasticity problem.

By symmetry, the location of drives and symmetry boundary conditions for solving the problem for each of the basic movements have symmetry. It should be noted that the types of symmetries for the basic movements of the "up-down" and "rotate in the same direction are the same."

Used indexes: $\upsilon$ - movement of disks up – down; $u$ - movement disks the left – right; $\omega$ - rotating of disks in the same direction.

To calculate the total energy is possible to consider only the first three basic movements, and the fourth can be neglected. Strain $\varepsilon_{ij}$ in the neck between the discs between the two disks with arbitrary displacement drive is a linear combination of base strains $\varepsilon_{ij}^{u0}, \varepsilon_{ij}^{\upsilon 0}, \varepsilon_{ij}^{\omega 0}$, namely:

$$\varepsilon_{ij} = \upsilon \varepsilon_{ij}^{\upsilon 0} + \omega \varepsilon_{ij}^{\omega 0} + u \varepsilon_{ij}^{u0}.$$

The corresponding energy

$$E = \frac{1}{2} c_{ijkl} (\upsilon \varepsilon_{ij}^{\upsilon 0} + \omega \varepsilon_{ij}^{\omega 0} + u \varepsilon_{ij}^{u0})(\upsilon \varepsilon_{kl}^{\upsilon 0} + \omega \varepsilon_{kl}^{\omega 0} + u \varepsilon_{kl}^{u0}),$$

where $c_{ijkl}$ - the elastic constants. Consider an isotropic elastic matrix, for which the non-zero elastic constants are [20]:

$$c_{1111} = c_{2222} = \lambda + 2\mu, \; c_{1122} = c_{2211} = \lambda, \; c_{1212} = c_{1221} = c_{2112} = c_{2121} = \mu,$$

$\lambda, \mu$ - Lame coefficients.

The total energy
$$E = E_1 + E_2 + E_3 + E_4$$

($E_i$ - energy region occupied $i$-th coordinate quarter) can be calculated as the sum of the energies in the quarters of the Cartesian coordinate system.

The solutions of problems for each type of basic movements have obvious symmetry (anti-symmetry) with respect to the coordinate axes. With regard to the symmetry of solutions

$$E_1 = \frac{1}{2} \int_0^\infty \int_0^\infty ((\lambda + 2\mu)((v\varepsilon_{11}^{v0} + \omega\varepsilon_{11}^{\omega 0} + u\varepsilon_{11}^{u0})^2 + (v\varepsilon_{22}^{v0} + \omega\varepsilon_{22}^{\omega 0} + u\varepsilon_{22}^{u0})^2)$$

$$+ 2\lambda(u\varepsilon_{11}^{u0} + v\varepsilon_{11}^{v0} + \omega\varepsilon_{11}^{\omega 0})(u\varepsilon_{22}^{u0} + \sigma\varepsilon_{22}^{v0} + \omega\varepsilon_{22}^{\omega 0}) + 4\mu(u\varepsilon_{12}^{u0} + v\varepsilon_{12}^{v0} + \omega\varepsilon_{12}^{\omega 0})^2) dxdy$$

$$E_2 = \frac{1}{2} \int_0^\infty \int_0^\infty ((\lambda + 2\mu)((v\varepsilon_{11}^{v0} - u\varepsilon_{11}^{u0} + \omega\varepsilon_{11}^{\omega 0})^2 + (v\varepsilon_{22}^{v0} - u\varepsilon_{22}^{u0} + \omega\varepsilon_{22}^{\omega 0})^2$$

$$+ 2\lambda(v\varepsilon_{11}^{v0} - u\varepsilon_{11}^{u0} + \omega\varepsilon_{11}^{\omega 0})(v\varepsilon_{22}^{v0} - u\varepsilon_{22}^{u0} + \omega\varepsilon_{22}^{\omega 0}) + 4\mu(v\varepsilon_{12}^{v0} - u\varepsilon_{12}^{u0} + \omega\varepsilon_{12}^{\omega 0})^2) dxdy$$

and (also accounting that due to the symmetry of the problem $E_3 = E_1$, $E_4 = E_2$). Then

$$E = 2(E_1 + E_2) = \int_0^\infty \int_0^\infty ((\lambda + 2\mu)((v\varepsilon_{11}^{v0} + \omega\varepsilon_{11}^{\omega 0} + u\varepsilon_{11}^{u0})^2 + (v\varepsilon_{22}^{v0} + \omega\varepsilon_{22}^{\omega 0} + u\varepsilon_{22}^{u0})^2 +$$

$$(v\varepsilon_{11}^{v0} - u\varepsilon_{11}^{u0} + \omega\varepsilon_{11}^{\omega 0})^2 + (v\varepsilon_{22}^{v0} - u\varepsilon_{22}^{u0} + \omega\varepsilon_{22}^{\omega 0})^2) +$$

$$2\lambda((v\varepsilon_{11}^{v0} + u\varepsilon_{11}^{u0} + \omega\varepsilon_{11}^{\omega 0})(v\varepsilon_{22}^{v0} + u\varepsilon_{22}^{u0} + \omega\varepsilon_{22}^{\omega 0}) + (v\varepsilon_{11}^{v0} - u\varepsilon_{11}^{u0} + \omega\varepsilon_{11}^{\omega 0})(v\varepsilon_{22}^{v0} - u\varepsilon_{22}^{u0} + \omega\varepsilon_{22}^{\omega 0}) +$$

$$4\mu((v\varepsilon_{12}^{v0} + u\varepsilon_{12}^{u0} + \omega\varepsilon_{12}^{\omega 0})^2 + (v\varepsilon_{12}^{v0} - u\varepsilon_{12}^{u0} + \omega\varepsilon_{12}^{\omega 0})^2) dxdy.$$

When squaring the sums in the integrands appear as squares strain and strain pair products. Given the symmetries of the integrals of certain pair products deformations cancel each other and the total energy is expressed by the integral over $W_1$ - the first quarter of the coordinate system

$$E = 2(\lambda + 2\mu) \iint_{W_1} ((v\varepsilon_{11}^{v0} + \omega\varepsilon_{11}^{\omega 0})^2 + (v\varepsilon_{22}^{v0} + \omega\varepsilon_{22}^{\omega 0})^2 + (u\varepsilon_{11}^{u0})^2 + (u\varepsilon_{22}^{u0})^2) dxdy +$$

$$4\lambda \iint_{W_1} ((v\varepsilon_{11}^{v0} + \omega\varepsilon_{11}^{\omega 0})(v\varepsilon_{22}^{v0} + \omega\varepsilon_{22}^{\omega 0}) + \varepsilon_{11}^{u0} \varepsilon_{22}^{u0}) dxdy +$$

$$8\mu \iint_{W_1} ((v\varepsilon_{12}^{v0} + \omega\varepsilon_{12}^{\omega 0})^2 + (u\varepsilon_{12}^{u0})^2) dxdy.$$

Expanding the brackets in (2) and introducing independent on $x, y$ multipliers $v^2, v\omega, \omega^2, u^2$ the integral sign, we obtain the following expression for the total energy of deformation of the matrix:

$$E = A\upsilon^2 + B\upsilon\omega + C\omega^2 + Du^2,$$

The coefficients of the quadratic form (3) are given by the following integrals

$$A = 2(\lambda + 2\mu)\iint_{W_1}((\varepsilon_{11}^{\upsilon 0})^2 + (\varepsilon_{22}^{\upsilon 0})^2)dxdy + 4\lambda\iint_{W_1}\varepsilon_{11}^{\upsilon 0}\varepsilon_{22}^{\upsilon 0}dxdy + 8\mu\iint_{W_1}(\varepsilon_{12}^{\upsilon 0})^2 dxdy,$$

$$B = 4(\lambda + 2\mu)\iint_{W_1}(\varepsilon_{11}^{\upsilon 0}\varepsilon_{11}^{\omega 0} + \varepsilon_{22}^{\upsilon 0}\varepsilon_{22}^{\omega 0})dxdy + 4\lambda\iint_{W_1}(\varepsilon_{11}^{\upsilon 0}\varepsilon_{22}^{\omega 0} + \varepsilon_{22}^{\upsilon 0}\varepsilon_{11}^{\omega 0})dxdy + 16\mu\iint_{W_1}\varepsilon_{12}^{\upsilon 0}\varepsilon_{12}^{\omega 0}dxdy,$$

$$C = 2(\lambda + 2\mu)\iint_{W_1}((\varepsilon_{11}^{\omega 0})^2 + (\varepsilon_{22}^{\omega 0})^2)dxdy + 4\lambda\iint_{W_1}\varepsilon_{11}^{\omega 0}\varepsilon_{22}^{\omega 0}dxdy + 8\mu\iint_{W_1}(\varepsilon_{12}^{\omega 0})^2 dxdy,$$

$$D = 2(\lambda + 2\mu)\iint_{W_1}((\varepsilon_{11}^{u 0})^2 + (\varepsilon_{22}^{u 0})^2)dxdy + 4\lambda\iint_{W_1}\varepsilon_{11}^{u 0}\varepsilon_{22}^{\upsilon 0}dxdy + 8\mu\iint_{W_1}(\varepsilon_{12}^{u 0})^2 dxdy.$$

Note that in the quadratic form (3), which expresses the energy of deformation of the matrix in terms of the mutual movement of discs, there is the off-diagonal term $B\upsilon\omega$, which reflects the movement of the "up-down" and rotating in the same direction are coupled. The presence of off-diagonal terms is a fundamental difference between vector problem of elasticity theory of scalar problems (heat conduction, electrostatics, etc.). The analogy with the scalar problem (for example, the problem of the electrical capacitance) takes place for the displacements "to the left to the right. To move "up and down" and "rotate in the same direction" there is no analog to scalar case, here the energy is the general quadratic form

$$A\upsilon^2 + B\upsilon\omega + C\omega^2.$$

This is due to the fact that the movement of the "up-down" and "rotation in one direction" generate in neck the local deformations of similar form, namely shear deformation.

The coefficients of the quadratic form (3) can be obtained by numerically solving the problems of the theory of elasticity for the 1st, 2nd and 3rd movements of basic drives and the subsequent numerical evaluation of integrals (4) (this can be done in the environment of ANSYS). As an example, the coefficients of the quadratic form (3) have been calculated for $\delta = 2\times 10^{-2}$, $R_2 = 1$. For these $\delta$ and $R_2$ the following values were obtained

$$A = 0.59\times 10^{-6}, B = 1.02\times 10^{-6}, C = 0.58\times 10^{-6}, D = 8.25\times 10^{-6}.$$

It is seen, the coefficient $B$ is not zero.

# 3. Scaling formulas for energy

The coefficients (5), as well as the harmonic (electrostatic) capacity of bodies depend only on the geometry of the inclusions. In the case under consideration include - discs. Therefore, the coefficients of (5) should depend only on the values $\delta$ and $R_2$. We show that in fact, the coefficients depend only on the ratio $\dfrac{\delta}{R_2}$, that is, the problem is self-similar to the spatial variables. Consider the transformation of coordinates

$$x_1 = xt, \; y_1 = yt \; .$$

The Dirichlet problem for the theory of elasticity of a self-similar, which makes the solution is not changed by the transformation (6). Consider integrals (5) giving the coefficients *A, B, C, D* of the quadratic form of energy (3). With the change of variables (6), the operators of differentiation are replaced in accordance with the formulas

$$\frac{\partial}{\partial x} = t\frac{\partial}{\partial x_1}, \; \frac{\partial}{\partial x} = t\frac{\partial}{\partial y_1} \; .$$

thus, in the integrals in (4) will be a factor $t^2$, at the same time at this change of variables the Jacobean

$$\frac{\partial(x,y)}{\partial(x_1,y_1)} = \frac{1}{t^2}$$

and by reducing these factors in the integrals (4) for change of variables (6) do not change, that is, they are invariants of the transformation (6). Then [21], the integrals in (4) (i.e., the coefficients A,B,C,D) depend only on the ratio $\dfrac{\delta}{R_2}$. For example, coefficients (5) are the same for any $\delta$ and $R_2$ whose ratio $\dfrac{\delta}{R_2} = 5 \times 10^{-2}$ .

We note that the considered problem describes the behavior of a pair of parallel fibers of circular cross-section in a plane perpendicular to its axis [22, 23].

## Notations

$R_1$ –radius large disk; $R_2$ – radius small disks; $\delta$ – the distance between small disks; $E_1$ – Young's modulus of the material matrix; $E_2$ – Young's modulus of the material disks; $v$ – Poisson's ratio; $u$ –movement by axes Ox; $\upsilon$ – movement by axes Oy; $u_r$ – radial displacement; $u_\varphi$ – angle displacement; $E$ –energy.

Indexes:

sh –neck; mat –matrix.